\documentclass[aip,jcp,amsmath,amssymb,reprint]{revtex4-2}
\usepackage{graphicx, float, xcolor, pgf}
\usepackage{physics, siunitx}
\usepackage[caption=false, justification=centerlast]{subfig}

\usepackage{hyperref}
\hypersetup{
  pdfauthor={Amartya Bose},
  pdftitle={Impact of solvent on state-to-state population transport in multistate systems using coherences},
  colorlinks=true,
  allcolors=.
}

\begin{document}
\title{Impact of solvent on state-to-state population transport in multistate systems using coherences}
\author{Amartya Bose}
\email{amartya.bose@tifr.res.in}
\affiliation{Department of Chemical Sciences, Tata Institute of Fundamental Research, Mumbai 400005, India}
\author{Peter L. Walters}
\email{peter.l.walters2@gmail.com}
\affiliation{Department of Chemistry and Biochemistry, George Mason University, Fairfax, Virginia 22030, USA}
\thanks{Both authors contributed equally to this work.}
\allowdisplaybreaks

\begin{abstract}
  Understanding the pathways taken by a quantum particle during a transport
  process is an enormous challenge. There are broadly two different aspects of
  the problem that affect the route taken. First is obviously the couplings
  between the various sites, which translates into the intrinsic ``strength'' of
  a state-to-state channel. Apart from the inter-state couplings, the solvents
  affecting the energies of the state, and their relative coupling strengths and
  time-scales form the second factor. This impact of dissipative media is
  significantly more difficult to analyze. Building on recently derived
  relations between coherences and population derivatives, we present an
  analysis of the transport that allows us to account for both the effects in a
  rigorous manner. We demonstrate the richness hidden behind the transport even
  for a relatively simple system, a 4-site coarse-grained model of the
  Fenna-Matthews-Olson complex. The effect of the local dissipative media is
  highly non-trivial. We show that while the impact on the total site population
  may be small, there are dramatic changes to the pathway taken by the transport
  process. The ability to untangle the dynamics at a greater granularity opens
  up possibilities in terms of design of novel systems with an eye towards
  quantum control.
\end{abstract}
\maketitle

\section{Introduction}\label{sec:intro}
Simulating complex chemical reactions in the condensed phase has been the holy grail
of computational and theoretical chemistry. This already difficult task becomes
even more arduous when the reaction involves the purely quantum mechanism of
tunneling. However, this is ubiquitous in various processes like exciton
transport in photosynthetic complexes, electron transfer, etc. In addition, many
exciton and electron transfer processes happen in extended systems where there
can be multiple pathways for the quantum ``particle'' to follow. A thorough
understanding of the contribution of these various pathways is necessary to
facilitate a more clear picture of the dynamics.

The simulation of the basic dynamics of quantum particles in a condensed phase
is quite challenging in and of itself. Approximations like Redfield and
F\"orster,~\cite{forsterZwischenmolekulareEnergiewanderungUnd1948} though often
used, are not universally applicable, especially in the strongly coupled regime.
For numerically exact simulations of dynamics of extended systems, approaches
based on tensor-networks have been gaining a lot of popularity. Most notable
among them are density matrix renormalization group
(DMRG)~\cite{whiteDensityMatrixFormulation1992,
  schollwockDensitymatrixRenormalizationGroup2005,
  schollwockDensitymatrixRenormalizationGroup2011a} and its time-dependent
variant.~\cite{whiteRealTimeEvolutionUsing2004} The family of multiconfiguration
time-dependent Hartree
(MCTDH)~\cite{beckMulticonfigurationTimedependentHartree2000,
  wangMultilayerFormulationMulticonfiguration2003} can also be thought of as being
based on tree tensor networks. However, the approaches often fail to account for
the effects of (a possible continuum of) translational and vibrational degrees
of freedom contributed by the solvent.

Methods based on simulating the reduced density matrix provide a lucrative
alternative to the above-mentioned methods for simulating these systems. Of
these, the quasi-adiabatic propagator path integral
(QuAPI)~\cite{makriTensorPropagatorIterativeI1995,
  makriTensorPropagatorIterativeII1995} and hierarchical equations of
motion (HEOM)~\cite{tanimuraTimeEvolutionQuantum1989,
  tanimuraReducedHierarchicalEquations2014, tanimuraNumericallyExactApproach2020}
are the most widely used. The development of small matrix
decomposition~\cite{makriSmallMatrixDisentanglement2020,
  makriSmallMatrixPath2020} of QuAPI has made it especially viable for
simulating large systems. Additionally, tensor networks have also been shown to
be exceptionally useful in increasing the efficiency of
path integral methods.~\cite{strathearnEfficientNonMarkovianQuantum2018,
  jorgensenExploitingCausalTensor2019, boseTensorNetworkRepresentation2021,
  bosePairwiseConnectedTensor2022} These tensor network-based ideas have very
recently been successfully extended to a multisite framework capable of
simulating the quantum dynamics of extended systems coupled with local
dissipative media.~\cite{boseMultisiteDecompositionTensor2022,
  boseEffectTemperatureGradient2022, boseTensorNetworkPath2022}

Studies of population dynamics conducted with these methods, while very rich in
information, are unable to provide a clear and unambiguous insight into the
mechanism of the transport. Consider an extended system with a non-trivial
topology allowing for long-ranged couplings between sites, and assume we are
interested in the transport of an exciton. For a given initial location of the
exciton, one would traditionally focus on the time-dependent population of the
exciton on each of the sites. We would have no further information on the route
or ``pathway'' that the exciton took to get to a particular site. Such
information, however, is crucial to optimization of materials for guided quantum
transport. An extremely na\"ive approach to analyzing the pathways would be to
track the route of the strongest couplings in the system Hamiltonian that leads
from the ``source'' to the ``sink.'' Such an approach would obviously miss out
on the effects of the dissipative media. A different approach has been recently
used to understand these pathways under a Lindbladian model Hamiltonian by
evaluating the transport of the base system vis-\`a-vis a system with a
particular chromophore dropped.~\cite{bakerRobustnessEfficiencyOptimality2015}
The idea is that dropping a chromophore that is a part of the primary pathways
would lead to a large decrease of transport efficiency.
Alternatively, \citet{wuEfficientEnergyTransfer2012} have used flux networks constructed on
integrated flux between sites and the flux balance
method~\cite{caoMichaelisMentenEquation2011} to analyze the pathways in the
Fenna-Matthews-Olson complex (FMO).

Recently, \citet{daniQuantumStatetoStateRates2022} have shown that the
instantaneous rate of change of the site population is related to the
off-diagonal terms of the reduced density matrix (also called the
``coherences'') and rigorously derived the rate constants specific to the
various state-to-state channels. While rates and kinetic models can often offer
deep insights,~\cite{ritzKineticsExcitationMigration2001,
  wuEfficientEnergyTransfer2012, caoMichaelisMentenEquation2011} in many ultrafast
chemical systems, the short-time dynamics, often called the transients, may be
very important. Rate theory generally fails for such processes and ones with
more than one primary time-scale. In these cases, it becomes crucial to shift
our attention from rates to the population transfer. ``Coherence maps'' are
visual representations of the time evolution of the off-diagonal terms of the
reduced density matrix and have recently been shown to capture important
features of the structure of the
Hamiltonian.~\cite{daniTimeEvolvingQuantumSuperpositions2022} Building on these
insights, we show how one can efficiently leverage the information in the
off-diagonal terms to understand the effect of dissipative media in modulating
the direct transport between sites as a function of time.

The paper is organized as follows. The analysis performed in this paper and its
connections with the previous works is outlined in Sec.~\ref{sec:method}.
Thereafter, we explore the excitation dynamics in a coarse-grained four-site
model of the Fenna-Matthews-Olson complex (FMO) with a focus on how this
information can be used in a directed manner to gain detailed insights into the
same. Finally, some concluding remarks and future outlook are presented in
Sec.~\ref{sec:conclusion}.

\section{Importance of Coherences in Direct Unmediated Population Transport}\label{sec:method}
Consider a system with $N$ sites or states coupled with arbitrary harmonic
baths. These baths may or may not be site-local. The Hamiltonian of such a
problem is generally of the form:
\begin{align}
  \hat{H} = \hat{H}_0 + \hat{H}_\text{SB}
\end{align}
where $\hat{H}_0$ is the Hamiltonian corresponding to the system  and
$\hat{H}_\text{SB}$ is the Hamiltonian corresponding to the system-bath
coupling. (It is assumed that the system is represented in a basis that
diagonalizes $\hat{H}_\text{SB}$.) Under Gaussian response, the harmonic baths
are often obtained from a simulation of the bath response
function.~\cite{makriLinearResponseApproximation1999,
  boseZerocostCorrectionsInfluence2022} Usually one simulates the time-dependent
population of each of the states. Here, we define the \emph{direct}
``state-to-state'' population transfer from state $k$ to state $j$ as the
population transfer between them without any intermediate state, also denoted by
$P_{j\leftarrow k}$. The objective is to be able to simulate $P_{j\leftarrow k}$
as a function of time.  Given that a ``pathway'' or ``route'' of transport is
nothing but a sequence of these state-to-state population transfers, it should
be possible to assemble a picture of the important pathways using them as the
building blocks.

If $P_j = \Tr\left(\tilde\rho(t) \dyad{j}\right)$, is the population of the
$j$th site, it is trivial to show that the time derivative of this population
can be expressed as:~\cite{daniQuantumStatetoStateRates2022}
\begin{align}
  \dv{P_j}{t}            & = \Tr\left[\tilde\rho(t) \hat{F}_j\right],\label{eq:deriv}  \\
  \text{where }\hat{F}_j & = \frac{i}{\hbar}\comm{\hat{H}_0}{\dyad{j}}.\label{eq:flux}
\end{align}
This commutator, $\hat{F}_j$, is exactly the same flux operator that is used for
rate theory.~\cite{millerQuantumMechanicalTransition1974,
  millerQuantumMechanicalRate1983} Though generally rate theory is formulated in
terms of the equilibrium correlation functions, it has been shown that the rate
for a two-state problem can be obtained as a ``plateau'' value of the
time-dependent non-equilibrium flux, Eq.~\ref{eq:deriv} and
\ref{eq:flux}, as well.~\cite{boseNonequilibriumReactiveFlux2017} However, here we are
not interested in a rate perspective. We rather want to understand the full time
dynamics with additional information about the channel-dependent contributions.
Following~\citet{daniQuantumStatetoStateRates2022}, we expand Eq.~\ref{eq:deriv}
to get
\begin{align}
  \dv{P_j}{t} & = \frac{i}{\hbar}\sum_k \mel{j}{\tilde\rho(t)}{k}\mel{k}{\hat{H}_0}{j} - \mel{j}{\hat{H}_0}{k}\mel{k}{\tilde\rho(t)}{j}.\label{eq:deriv_expand}
\end{align}
For a real symmetric time-independent system Hamiltonian,
\begin{align}
  \dv{P_j}{t} & = -\frac{2}{\hbar}\sum_k\mel{j}{\hat{H}_0}{k} \Im\mel{j}{\tilde\rho(t)}{k}\label{eq:deriv_symm},
\end{align}
Equations~\ref{eq:deriv_expand} and~\ref{eq:deriv_symm} can be interpreted in
terms of the rates along the different state-to-state
channels.~\cite{daniQuantumStatetoStateRates2022} By comparing
Eq.~\ref{eq:deriv_symm} with Eq.~\ref{eq:deriv_expand}, one can, for any $k$,
interpret the term with $\Im\mel{k}{\tilde\rho(t)}{j}$ as the rate of flow from
site $k$ into $j$ and the term with $\Im\mel{j}{\tilde\rho(t)}{k}$ as the rate
of flow from site $j$ to $k$. The time evolution of the individual coherences,
$\mel{j}{\tilde\rho(t)}{k}$, as captured through coherence maps, also shows very
interesting features reflecting the system dynamics and
equilibrium.~\cite{daniTimeEvolvingQuantumSuperpositions2022}

Often the full population dynamics is a fruitful interrogative tool for
understanding the system. When it comes to ideas of quantum design and
understanding pathways of the excitation energy transport (EET) processes, it
seems to be helpful to think in terms of site-to-site population transfer.
The coherences allow us a crucial ability to express the direct and unmediated
transfer of population between different sites. One can directly use
Eq.~\ref{eq:deriv_expand} to partition the total population change at a site
into the contributions from each state-to-state channel. We define the
time-dependent population flow from the $k$th site to the $j$th site as
\begin{align}
  P_{j\leftarrow k}(t) & = \frac{i}{\hbar}\int_0^t \dd{t'}\left(\mel{j}{\tilde\rho(t')}{k}\mel{k}{\hat{H}_0}{j}\right.\nonumber \\
                       & \left.- \mel{j}{\hat{H}_0}{k}\mel{k}{\tilde\rho(t')}{j}\right)\label{eq:cdpt}.
\end{align}
For a real symmetric time-independent system Hamiltonian, using
Eq.~\ref{eq:deriv_symm}, this reduces to
\begin{align}
  P_{j\leftarrow k}(t) & = -\frac{2}{\hbar}\mel{j}{\hat{H}_0}{k} \int_0^t \dd{t'} \Im\mel{j}{\tilde\rho(t')}{k}\label{eq:cdpt_simplified}.
\end{align}
Notice that the state-to-state population flow between two sites is proportional
to the Hamiltonian matrix element between them. This is what we would have
na\"ively concluded. However, the proportionality constant is related to the
time integral of the coherence. This captures the solvent effect on the system
dynamics. These state-to-state populations are independent of how the simulation
was done, and therefore can be applied to any level of simulation as desired.
There are no further approximations over and above the ones used for simulating
the time-dependent reduced density matrix of the system. Notice that
Eqs.~\ref{eq:cdpt} and~\ref{eq:cdpt_simplified} uphold detailed balance in the
sense that $P_{j\leftarrow k}(t) = -P_{k\leftarrow j}(t)$ and that
$P_{j\leftarrow j}(t)=0$ for all $j$, encoding the fact that there cannot be any
population transfer from a site to itself. We would like to point out that the
long time limits of the time-dependent state-to-state population flow is the
same as the quantum integrated population fluxes derived
by~\citet{wuEfficientEnergyTransfer2012} and used to construct the flux
networks.

Finally, the time dependent population of the $j$th site can be expressed as
\begin{align}
  P_j(t) & = P_j(0) + \sum_{k\ne j}P_{j\leftarrow k}(t).
\end{align}
The ability to partition the time-dependent population on a site into the
components along various channels is important for understanding the effects
resulting from the non-trivial interactions between specific changes in the
dissipative media and the system Hamiltonian. If the system is thought of as a
graph, with the sites being the vertices and the edges being the various
inter-site connections, then the time-integrated coherences decompose the
time-evolution of the population on a site (vertex) along all the edges that are
incident on it.

\section{Results}\label{sec:results}

\begin{figure}
  \centering
  \includegraphics[scale=0.5]{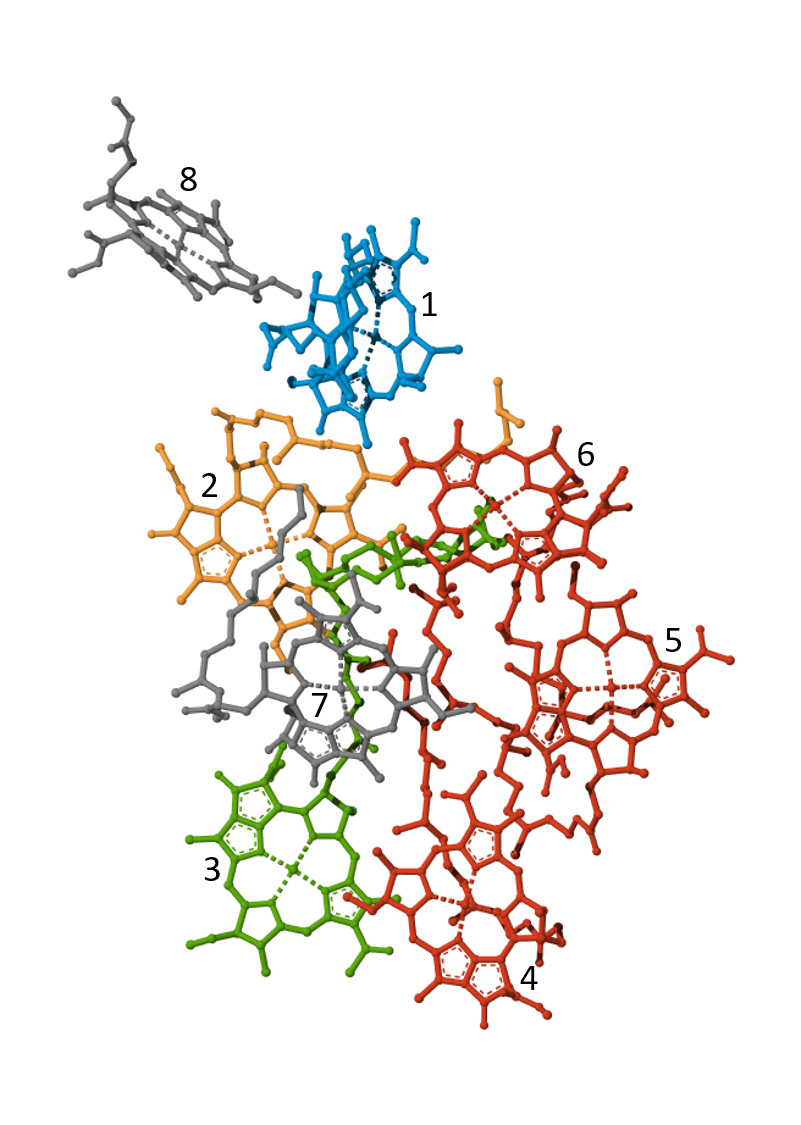}
  \caption{Fenna-Matthews-Olson complex with the bacteriochlorophyll units
    colored by the coarse-grained units used. Blue: Coarse-grained site 1.
    Orange: Coarse-grained site 2. Green: Coarse-grained site 3. Red:
    Coarse-grained site 4. Gray: Ignored.}\label{fig:FMO}
\end{figure}

\begin{figure}
  \subfloat[Site-dependent spectral densities]{\includegraphics{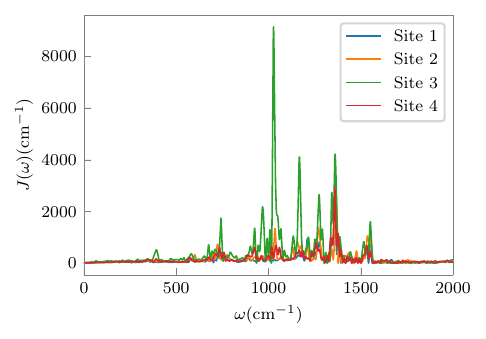}}
  
  \subfloat[Average spectral density]{\includegraphics{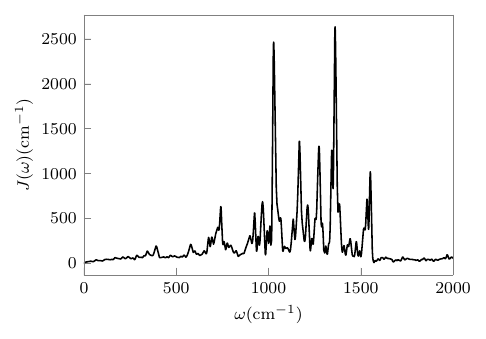}}
  \caption{Site-dependent and average spectral densities for the first four
    bacteriochlorophyll units in FMO obtained
    from~\citet{maityDFTBMMMolecular2020}}\label{fig:Jw}
\end{figure}

To demonstrate the utility of this analysis of the state-to-state population
transfers leveraging the information of the coherences, consider a
coarse-grained system modeled on the FMO complex. FMO is a naturally occurring
light-harvesting complex with eight bacteriochlorophyll monomeric sites. It is
ubiquitous as a model for excitonic transport and provides a very rich set of
dynamical features owing to the non-linear inter-site couplings. To enable a
thorough exploration of the impact of the vibrational modes on the transfers
through various state-to-state channels, we simplify the system by
coarse-graining it to include the four most relevant sites. For FMO, it is known
that if bacteriochlorophyll site 1 is initially excited, the primary pathway is
$1\rightarrow 2\rightarrow 3$ and the secondary pathway leads from $1\rightarrow
  6\rightarrow 5\rightarrow 4\rightarrow 3$. Thus, in our coarse-grained model we
keep sites 1, 2, and 3 as is, reduce sites 4, 5, and 6 into a new renormalized
4th site and omit sites 7 and 8 entirely.  This is shown in Fig.~\ref{fig:FMO}.
Similar to the full FMO, we expect the model to have a primary pathway of
$1\rightarrow 2\rightarrow 3$ and a secondary pathway of $1\rightarrow
  4\rightarrow 3$.

The coarse-grained FMO model along with its interactions with the local
vibrational baths is described by the following Hamiltonian:
\begin{align}
  \hat{H}           & = \hat{H}_0 + \hat{H}_\text{SB},                                                                                                                                      \\
  \hat{H}_0         & = \sum_{k=1}^4 \epsilon_k\dyad{k} + \sum_{j\ne k} h_{j,k}\dyad{j}{k},                                                                                                 \\
  \hat{H}_\text{SB} & = \sum_{k=1}^4 \sum_{j=1}^{N_\text{osc}} \frac{p_{kj}^2}{2m_{kj}} + \frac{1}{2}m_{kj}\omega_{kj}^2\left(x_{kj} - \frac{c_{kj}\dyad{k}}{m_{kj}\omega_{kj}^2}\right)^2,
\end{align}
where $\omega_{kj}$ and $c_{kj}$ are the frequency and coupling of the $j$th
harmonic mode of the bath corresponding to the $k$th site. The electronic
excitation energies are given by $\epsilon_k$ and the inter-site couplings are
given by $h_{j,k}$.

The frequencies and couplings of the baths are characterized by the spectral
density defined as
\begin{align}
  J_k(\omega) & = \frac{\pi}{2}\sum_j \frac{c_{kj}^2}{m_{kj}\omega_{kj}}\delta(\omega_{kj} - \omega).
\end{align}
This can be calculated as the Fourier transform of the energy-gap
autocorrelation function simulated using molecular dynamics. The site-dependent
spectral densities and Hamiltonian for FMO have been recently calculated
by~\citet{maityDFTBMMMolecular2020} We use these parameters as
the starting point for our exploration. The Hamiltonian corresponding to this
coarse-grained model is given in the supplementary information. It is based on
the FMO Hamiltonian obtained by~\citet{maityDFTBMMMolecular2020} In the original
work the simulation was done using QM/MM MD trajectories, TD-LC-DFTB was used
for the site energies and TrESP for the
couplings.~\cite{maityDFTBMMMolecular2020} The average and the site-dependent
spectral densities are shown in Fig.~\ref{fig:Jw} for reference. In our
exploration of the FMO model, we will change the spectral densities in various
ways that shall be described. However, the parameters for the system Hamiltonian
will always remain the same to ensure that the effects that we see arise solely
out of the vibrational baths.

\begin{figure}
  \centering
  \includegraphics{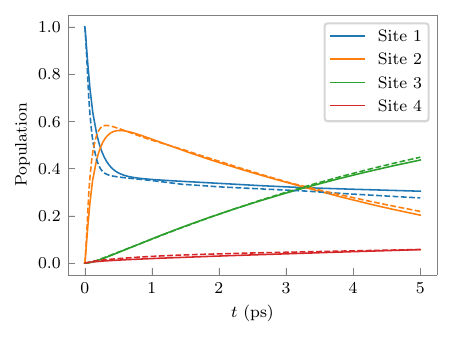}
  \caption{Excitonic population on different sites as a function of time. Solid line: Average spectral density. Dashed line: Different spectral densities.}\label{fig:popln_base}
\end{figure}
\begin{figure}
  \centering
  \hspace*{-0.5cm}
  \subfloat[Site 1]{\includegraphics{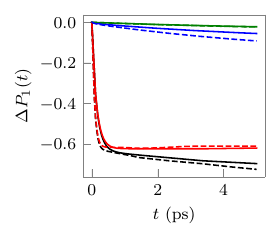}}
  ~\subfloat[Site 2]{\includegraphics{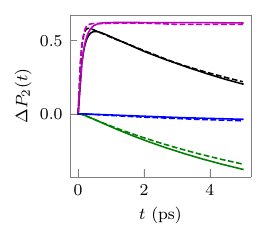}}
  
  \hspace*{-0.5cm}
  \subfloat[Site 3]{\includegraphics{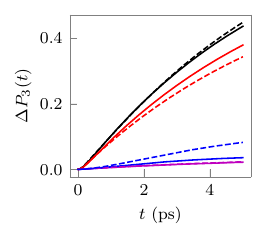}}
  ~\subfloat[Site 4]{\includegraphics{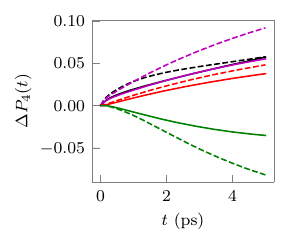}}
  \caption{Transfer pathways of excitonic population corresponding to each site
    with different spectral densities. Solid line: average bath. Dashed line:
    Different spectral densities. Black: Total change of population of the site.
    Magenta: Change due to site 1 ($P_{\star\leftarrow 1}(t)$). Red: Change due
    to site 2 ($P_{\star\leftarrow 2}(t)$). Green: Change due to site 3
    ($P_{\star\leftarrow 3}(t)$). Blue: Change due to site 4
    ($P_{\star\leftarrow 4}(t)$).}\label{fig:cdpt_base}
\end{figure}
Figure~\ref{fig:popln_base} shows the excitonic population corresponding to each
of the sites for the site-specific and average spectral densities. (This
information can, in principle, be calculated using many methods. Here the
simulations have been conducted using the tensor network path integral
method~\cite{boseTensorNetworkRepresentation2021} based on Feynman-Vernon
influence functional.) We notice that changing the average spectral density to
the site-specific spectral densities has minor effects on the dynamics of
bacteriochlorophyll sites 1 and 2 and negligible effects on the populations of
sites 3 and 4. A key drawback of this population picture is that it washes away
a lot of details. At this level, one cannot answer questions such as how does
the transfer from site 1 to site 2, $P_{2\leftarrow 1}(t)$, change in switching
between the two descriptions. Or what happens to the various contributions to
the site 3 population?

The analysis of the imaginary part of coherences allows us to answer these
questions. In Fig.~\ref{fig:cdpt_base}, we show the population dynamics of
specific sites along with the individual contributions. The first thing that one
immediately observes is that the primary flow of excitonic population happens
along $1\rightarrow 2\rightarrow 3$. To see this consider that the excitation
starts on site 1. The biggest transport happens from 1 to 2 in
Fig.~\ref{fig:cdpt_base}~(a) (red line). Then looking at where the population
goes from site 2, we see that the maximum amount goes to $3$ in
Fig.~\ref{fig:cdpt_base}~(b) (green line). By a similar analysis, we find a
secondary, slower, pathway that leads from site 1 to site 3 via site 4
($1\rightarrow 4\rightarrow 3$).  Additionally, one sees a non-insignificant
contribution from $1\rightarrow 2\rightarrow 4\rightarrow 3$. The direct
transfer from site 1 to site 3 is the least important of these. While the
ability to analyze the primary pathways immediately is obvious from
Fig.~\ref{fig:cdpt_base}, we would like to emphasize the power of the method in
terms of disqualifying unimportant pathways as well.  Notice that though site 1
transfers population into site 4, site 4 only transfers population into site 3.
Therefore, a path like $1\rightarrow 4\rightarrow 2\rightarrow 3$ is not
important.

We notice that with the site-specific spectral densities the excitonic
flow along $1\rightarrow 2\rightarrow 3$ is decreased coupled with an increased
flow along the $1\rightarrow 4\rightarrow 3$ pathway. As for the other two
pathways, the flow along $1\rightarrow 2\rightarrow 4\rightarrow 3$ increases,
and the direct transfer $1\rightarrow 3$ remains the same. These changes in the
exitonic pathways are evidenced by the fact that in going from the average to
the site specific spectral densities, the direct transfer from site 2 to site 3
(red curve in Fig.~\ref{fig:cdpt_base}~(c)) shows a decrease and the transfer
from 4 to 3 (blue curve in the same figure) shows an increase. Furthermore,
while there is an increase in both the transfer from site 1 to site 4 (magenta
curve in Fig.~\ref{fig:cdpt_base}~(d)) and site 2 to site 4 (red curve in the
same figure), the increase in $1\rightarrow 4$ is much larger. An explanation
for these changes can be made by looking at $P_{3\leftarrow 2}(t)$ (red curve in
Fig.~\ref{fig:cdpt_base}~(c)). We notice that for the site-specific spectral
densities, the direct transfer from site 2 to 3 seems to be somewhat restricted
causing a rerouting of the excitation through site 4.

\begin{figure}
  \subfloat[Scan on site 3]{\includegraphics{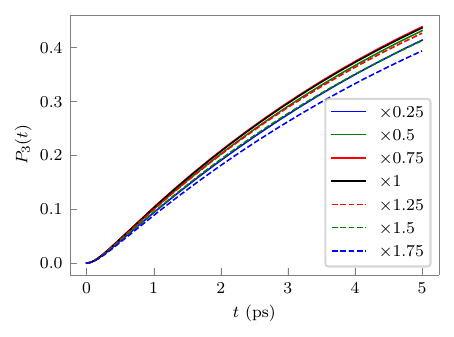}}
  
  \subfloat[Scan on site 2]{\includegraphics{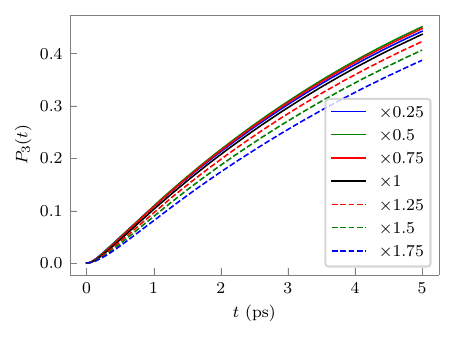}}
  \caption{Total population of site 3 as a function of time on scanning the
    reorganization energies on site 3 and site 2 respectively.}\label{fig:pop_scan}
\end{figure}

\begin{figure*}
  \subfloat[Scan on 3: $1\rightarrow 2$]{\includegraphics{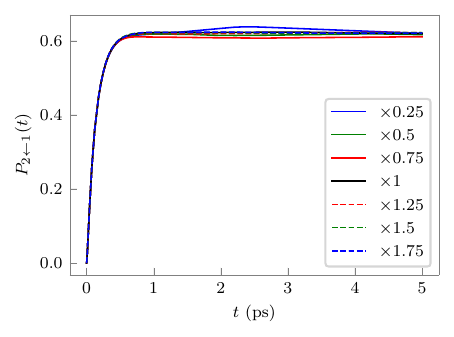}}
  ~\subfloat[Scan on 3: $2\rightarrow 3$]{\includegraphics{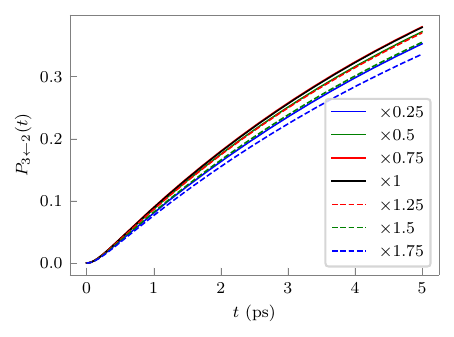}}
  
  \subfloat[Scan on 2: $1\rightarrow 2$]{\includegraphics{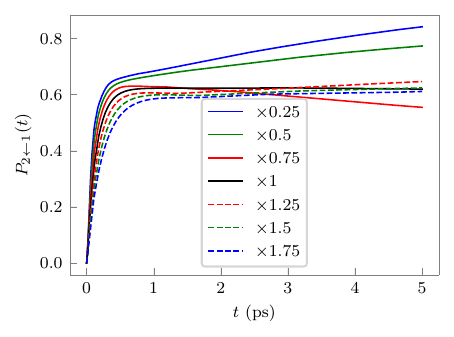}}
  ~\subfloat[Scan on 2: $2\rightarrow 3$]{\includegraphics{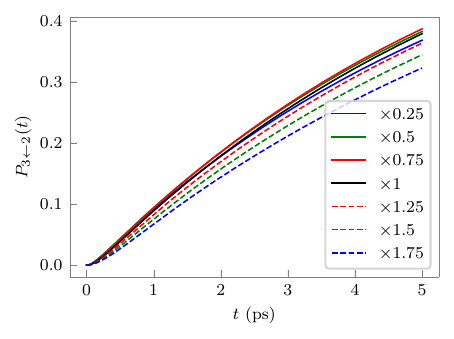}}
  \caption{Population transfer along $1\rightarrow 2$ and $2\rightarrow 3$ for
    components of the primary pathway when the site-specific reorganization
    energies on site 3 and site 2 are scanned.}\label{fig:path123_scan}
\end{figure*}

\begin{figure*}
  \subfloat[Scan on 3: $1\rightarrow 4$]{\includegraphics{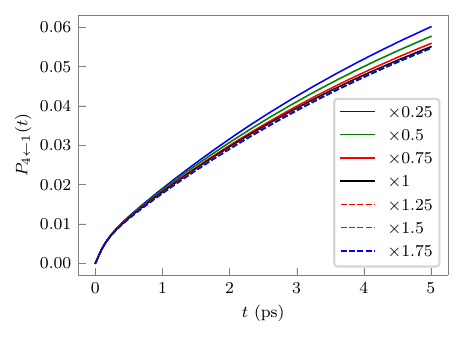}}
  ~\subfloat[Scan on 3: $4\rightarrow 3$]{\includegraphics{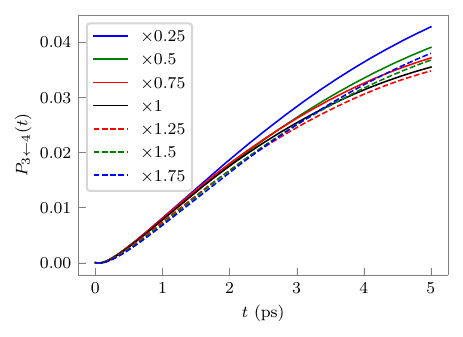}}
  
  \subfloat[Scan on 2: $1\rightarrow 4$]{\includegraphics{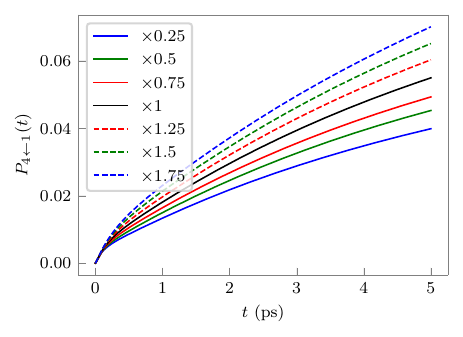}}
  ~\subfloat[Scan on 2: $4\rightarrow 3$]{\includegraphics{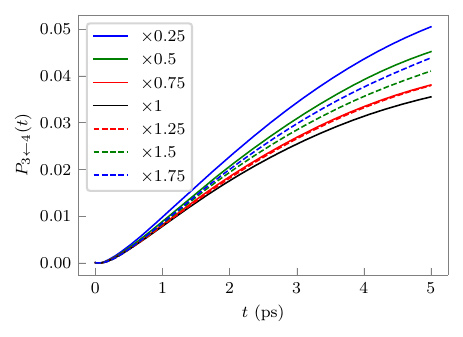}}
  \caption{Population transfer along $1\rightarrow 4$ and $4\rightarrow 3$ for
    components of the primary pathway when the site-specific reorganization
    energies on site 3 and site 2 are scanned.}\label{fig:path143_scan}
\end{figure*}

Apart from this rather broad overview of the pathways of excitation dynamics, a
state-to-state analysis can uncover a wide variety of other features. For example, we can
determine that the actual direct transfer from site 1 to site 2, $P_{2\leftarrow
      1}(t)$, stops after around \SI{0.5}{ps}. Thus, the red line in
Fig.~\ref{fig:cdpt_base}~(a) and the magenta line in
Fig.~\ref{fig:cdpt_base}~(b) becomes practically flat around that time. This is
the case even though the populations of both sites 1 and 2 keep changing
throughout the period of simulation. Additionally, on careful observation of the
plots Figs.~\ref{fig:cdpt_base}~(a) and (b), it is seen that around \SI{3}{ps}
-- \SI{4}{ps}, there is a backflow of population from site 2 to site 1 in case
of different baths. (Notice the small bump in the dashed red line in
Fig.~\ref{fig:cdpt_base}~(a) around that point.) Neither of these two analysis
would have been evident from a rate theory perspective.

To explore the effect of the site-dependence of the spectral density, we
systematically change the reorganization energies on single sites using the
average bath as a starting point. We scale the reorganization energies on site 3
and site 2 with factors ranging from 0.25 and 1.75 in steps of 0.25.
Figure~\ref{fig:pop_scan} shows the change in the population dynamics of site 3.
Notice that in both cases, the population curve initially increases but then
starts to decrease, though the reorganization energy where the maximum transfer
occurs is different in the two cases. This behavior is similar to the inverted
region in Marcus theory of electron
transfer.~\cite{marcusChemicalElectrochemicalElectronTransfer1964} It is
interesting that although the maximum transfer is obtained at different values,
at their respective maximum reorganization energies, the two curves look
remarkably similar. However, this apparent similarity hides differences in the
mechanism.

Let us consider the two pathways --- the primary one, $1\rightarrow 2\rightarrow
  3$ and the secondary one, $1\rightarrow 4\rightarrow 3$ separately.
Figure~\ref{fig:path123_scan} demonstrates the changes on the population
dynamics along the primary pathway. In Fig.~\ref{fig:path123_scan}~(b) and (d),
we see that the curves are very similar to Fig.~\ref{fig:pop_scan}~(a) and (b).
This implies that the transfer $2\rightarrow 3$ accounts for the main parts of
the dynamics of the excitation population on site 3. The transfer from site 2 to
site 3, in Fig.~\ref{fig:path123_scan}~(b) and (d), seems to hit a maximum and
decrease as we go away from it. This behavior is similar to the inverted region
of Marcus theory of rate of electron transfer. Interestingly, the transfer from
$1\rightarrow 2$ seems to plateau and vary slightly around a constant value.
However, things become completely different when the reorganization energy on
site 2 is scanned (Figs.~\ref{fig:path123_scan}~(c) and~(d)). There is
apparently very little pattern to the transfer from site 1 to site 2 as seen in
Fig.~\ref{fig:path123_scan}~(c). The lack of a Marcus-like inversion region is
because of the relatively large coupling between sites 1 and 2 which breaks
perturbation theory. Additionally, in all these simulations, one finds that the
transfer from site 1 to site 2 happens at the smallest time-scales. There are
further counter-intuitive features of the state-to-state population transfer
between sites 1 and 2. At a scaling of 0.75, there is an initial transfer from
site 1 to site 2, but around $\SI{0.5}{\ps}$ the transfer reverses and
population starts moving from 2 to 1. At a scaling of 1.25, there seems to be a
backflow around $\SI{1}{\ps}$ from site 2 to 1, and then the regular flow
resumes. For a scaling of 0.25, 0.5, 1.5 and 1.75, there are clearly two
different time scales involved, a faster one upto around $\SI{0.5}{\ps}$ and a
slower one after that. The transfers seem to be linear rather than exponential
in these two domains. These behaviors are only observable because of the ability
to partition the population transfer on a site-to-site channel basis.

Finally, let us turn our attention to the secondary pathway of $1\rightarrow
  4\rightarrow 3$. The impact of the scans on the two direct transfers that make
up this pathway are shown in Fig.~\ref{fig:path143_scan}. First notice that on
scanning the reorganization energy on site 3 (Fig.~\ref{fig:path143_scan}~(a)
and~(b)), both the direct transfers seem to decrease with the reorganization
energy and then stop changing. When scanning the reorganization energy on site
2, the transfer from $1\rightarrow 4$ seems to increase monotonically.  However,
the transfer from $4\rightarrow 3$ shows a decrease followed by an increase,
resulting in a minimum of amount of transfer.  These changes caused in the
$1\rightarrow 4\rightarrow 3$ pathway is very surprising, given that the
reorganization energy on site 2, which is not a part of this pathway, is being
scanned. (The data corresponding to the other channels, though not explored
here, has been shown in the supplementary information for completeness.)

\section{Conclusion}\label{sec:conclusion}
Many exact and approximate methods exist that can simulate dynamics in complex
systems coupled with solvents and vibrational modes. However, it is a
significantly different and more difficult challenge to understand the exact
routes that the transport process takes. Na\"ive approaches of looking at the
inter-site couplings fall short because of their failure to account for the
non-trivial effects of the solvent modes. In this paper, we have presented a
novel technique for analyzing the dynamics that yields the contribution of each
channel.

There has been a recent realization of the importance of the coherences or
off-diagonal terms in understanding the dynamics. It has been shown that the
time-derivative of the site populations can be written as a linear combination
of the imaginary part of the coherences.~\cite{daniQuantumStatetoStateRates2022}
The dynamics of these coherences have also been explored quite thoroughly under
different conditions.~\cite{daniTimeEvolvingQuantumSuperpositions2022} Based on
the relation of the time-derivatives with the imaginary parts of the coherences,
we show that the change in the population of a site can be trivially decomposed
into the contributions coming from different channels. Thus one can, using the
coherences, study the effects of the solvent and temperature on the direct and
unmediated transport between any pair of sites.

Employing this insight, we can start to untangle the dynamics of systems with complex
inter-site couplings. As demonstrated in the 4-site model based on the FMO, the
insights uncovered can often be very non-trivial. From a fundamental
perspective, just because the total population on a site has a relatively
regular pattern, it is not necessary that the regularity is there in all the
individual contributions. Similarly, the total population showing some strange
feature does not imply the existence of the weirdness in each of the
contributory dynamics. What is possibly equally important to a fundamental
understanding is the fact that changing the vibrational profile on a single site
not only affects the pathways involving that site, but other pathways as well.
This has important implications in trying to design materials and engineer
specific outcomes in complex open quantum systems.

An analysis of the coherences reveals a wealth of information that lay hidden in
the dynamics of the reduced density matrix. Extending the explorations in
Ref.~\cite{daniTimeEvolvingQuantumSuperpositions2022}, it is now possible to
asssociate causes with the various changes that happen in the total population
dynamics. In the near future, we will utilize these ideas in understanding other
processes beyond exciton transport, especially complex reactions with multiple
pathways like proton-coupled electron transfer and multi-proton transfers.
Finally, the fact that the various analyses of coherence done here and earlier
elsewhere~\cite{daniQuantumStatetoStateRates2022,
  daniTimeEvolvingQuantumSuperpositions2022} are not dependent on any single
method of simulation of the time evolved reduced density matrix makes
these ideas universally applicable.

\section*{Acknowledgment}
We acknowledge Nancy Makri for useful discussions and inputs.

\bibliography{library.bib}
\end{document}